\documentclass[conference]{IEEEtran}

\usepackage{amsmath}
\usepackage{amssymb}

\usepackage{mdframed}
\usepackage{tikz}
\usepackage{booktabs}
\begin{document}

\title{Scalable Multi-domain Trust Infrastructures for Segmented Networks}

\author{
    \IEEEauthorblockN{
        Sam Grierson\IEEEauthorrefmark{1},
        William J Buchanan\IEEEauthorrefmark{1},
        Craig Thomson\IEEEauthorrefmark{1},
        Baraq Ghaleb\IEEEauthorrefmark{1},
        Leandros Maglaras\IEEEauthorrefmark{1},
        Chris Eckl\IEEEauthorrefmark{2},
    }\\
    \IEEEauthorblockA{\IEEEauthorrefmark{1} 
        Blockpass ID Lab, Edinburgh Napier University, Edinburgh, UK\\
        Email: \{s.grierson2, b.buchanan, c.thomson3, b.ghaleb, l.maglaras\}@napier.ac.uk
    }
    \IEEEauthorblockA{\IEEEauthorrefmark{2} 
        Condatis Group Limited, Edinburgh, UK\\
        Email: chris.eckl@condatis.com
    }
}

\maketitle

\begin{abstract}
Within a trust infrastructure, a private key is often used to digitally sign a transaction, which can be verified with an associated public key. Using PKI (Public Key Infrastructure), a trusted entity can produce a digital signature, verifying the authenticity of the public key. However, what happens when external entities are not trusted to verify the public key or in cases where there is no Internet connection within an isolated or autonomously acting collection of devices? For this, a trusted entity can be elected to generate a key pair and then split the private key amongst trusted devices. Each node can then sign part of the transaction using their split of the shared secret. The aggregated signature can then define agreement on a consensus within the infrastructure. Unfortunately, this process has two significant problems. The first is when no trusted node can act as a dealer of the shares. The second is the difficulty of scaling the digital signature scheme. This paper outlines a method of creating a leaderless approach to defining trust domains to overcome weaknesses in the scaling of the elliptic curve digital signature algorithm. Instead, it proposes the usage of the Edwards curve digital signature algorithm for the definition of multiple trust zones. The paper shows that the computational overhead of the distributed key generation phase increases with the number of nodes in the trust domain but that the distributed signing has a relatively constant computational overhead.
\end{abstract}

\begin{IEEEkeywords}
DKG, ECDSA, EdDSA, Trust infrastructures
\end{IEEEkeywords}

\section{Introduction}
Within critical infrastructure protection, there are often a large number of devices which need to intercommunicate and gather and process data. Each node can generate a key pair and identify themselves with a private key to enhance trust. Using these generated keys, overlay networks may need to be created in which devices could be part of multiple trust domains.

As networked infrastructures scale, they often still employ a client-server approach in which a leader is defined for the control and organisation of an infrastructure. Nonetheless, this situation can pose challenges if the leader acts maliciously or becomes unavailable, raising the need for a more resilient trust-based infrastructure. In this resilient infrastructure,  nodes should have the capacity to self-organize while maintaining security through the use of distributed digital signature cryptography, even in the absence of a central leader. While this setup can enable a leaderless infrastructure, it raises the issue of multiple trust zones across the network and how specific nodes will bind with a distributed private key and associated public key. Unfortunately, many existing methods use the elliptic curve digital signature algorithm, which does not scale well for signature aggregation. Hence, this paper outlines a new distributed trust framework based on the Edwards-curve digital signature algorithm that can scale trusted infrastructures.

This paper aims to build on the usage of the Edwards curve digital signature algorithm with the signature aggregation method proposed by Komlo et al. \cite{komlo2021frost} and propose a framework which allows for the creation of multiple trust zones where there are no leaders, and where no private keys ever have to be revealed. The core contributions are reviewing existing methods and defining a multi-zone trust domain without a leader.

\section{Preliminaries}

\subsection{Elliptic Curve Cryptography}

Proposed independently in the late 80s by Koblitz \cite{koblitz1987} and Miller \cite{miller1986}, Elliptic Curve Cryptography (ECC) has quickly become the preference for establishing public-key cryptosystems. The rapid uptake in ECC-based cryptosystems results from the smaller key sizes and more efficient implementations than their non-ECC counterparts, such as the Digital Signature Standard (DSS) \cite{kravitz1993}.

Several elliptic curves are used in practical applications. The curves themselves are typically denoted in the Weierstrass form $E : y^2 = x^3 + ax + b$ where $a$ and $b$ are elements of a finite field $\mathbb{F}_p$ for a small prime $p > 3$. For a curve $E(\mathbb{F}_p)$ a cryptographic protocol uses a subgroup of $(x, y) \in \mathbb{F}^2_p$ solutions to the equation of the curve, plus the point at infinity. The size of this cryptographic subgroup is denoted by $\lvert E(\mathbb{F}_p) \rvert$, and the prime order is denoted by $n$. A fixed generator point in the cyclic subgroup is denoted by $G \in E(\mathbb{F}_p)$.

Two standardised algorithms utilise ECC in their construction: The Elliptic Curve Digital Signature Algorithm (ECDSA) \cite{johnson2001} and the Edwards-curve Digital Signature Algorithm (EdDSA) \cite{josefsson2017}.  

ECDSA was proposed as a variant of DSA using ECC by Johnson, \emph{et al.} \cite{johnson2001}, and has subsequently been standardised by NIST in the FIPS 186-4 \cite{barker2013} and FIPS 186-5 \cite{moody2023} standards. ECDSA requires the definition of both a conversion function $\textsf{conv} : E(\mathbb{Z}_p) \rightarrow \mathbb{Z}_p$ for converting elliptic curve subgroup elements into integers and a collision-resistant hash function $\textsf{H} : \{0, 1\}^\ast \rightarrow \mathbb{Z}_p$.

ECDSA is formally defined as a set of three algorithms. The first is the key generation algorithm $\textsf{ECDSA}.\textsf{gen}(1^\lambda)$ which randomly generates a private key $\textsf{sk} \in \mathbb{Z}_p$ of bit length $\lambda$ and computes the public key $\textsf{pk} := \textsf{sk} \cdot G \in E(\mathbb{Z}_p)$.

The second is the signing algorithm $\textsf{ECDSA}.\textsf{sign}_{\textsf{sk}}(m)$ which takes a message $m \in \{0, 1\}^\ast$ and performs the following steps:
\vspace{5pt}
\begin{enumerate}
    \item Compute $h := \textsf{H}(m) \in \mathbb{Z}_p$.
    \item Uniformly sample a $k \in \mathbb{Z}^\ast_p$ and compute $r := \textsf{conv}(k \cdot G) \in \mathbb{Z}_p$.
    \item Compute $s := k^{-1} \cdot (h + r \cdot \textsf{sk}) \in \mathbb{Z}_p$
    \item return $\sigma := (r, s)$
\end{enumerate}

The third is the verification algorithm $\textsf{ECDSA}.\textsf{vrfy}_{\textsf{pk}}(\sigma, m)$ which, given the signature $\sigma$ and the message $m$ performs the following steps:
\vspace{5pt}
\begin{enumerate}
    \item Compute $h := \textsf{H}(m) \in \mathbb{Z}_p$.
    \item Compute $R := G \cdot \sigma^{-1} \cdot h + \textsf{pk} \cdot \sigma^{-1} \cdot r$
    \item If $R \neq 1$ and $\textsf{conv}(R) = r$ return \texttt{accept} else return \texttt{reject}.
\end{enumerate}

Due to some of the problems related to the practical implementation of ECDSA, EdDSA was proposed by Bernstein \emph{et al.} \cite{bernstein2012} and subsequently standardised in RFC 8032 \cite{josefsson2017} and NIST's FIPS 186-5 \cite{moody2023}. While still being an ECC-based signature scheme, EdDSA uses a twisted Edwards curve defined by the equation $E : ax^2 + y^2 = 1 + bx^2y^2$ over a finite field $\mathbb{F}_p$ for small prime $p > 3$. Similar to ECDSA, EdDSA requires the definition of a collision-resistant hash function $\textsf{H} : \{0, 1\}^b \rightarrow \{0, 1\}^{2b}$,  but EdDSA differs in that its hash function takes a bit string of length $b$ and outputs a bit string of length $2b$. The EdDSA signature scheme then uses a Fiat-Shamir transformed Schnorr-like identification protocol \cite{schnorr1990} to generate the cryptographic signature.

For the following, assume that bit strings are interpreted as elements in $\mathbb{Z}_p$ when unspecified. EdDSA, much like ECDSA, is defined as a set of three algorithms. The first is the key generation algorithm $\textsf{EdDSA}.\textsf{gen}(1^\lambda)$ which randomly generates the private key $\textsf{sk} \in \{0, 1\}^\lambda$ and compute the public key $\textsf{pk} := k \cdot G \in E(\mathbb{Z}_p)$ where $k$ is the first $\lambda$ bits of $\textsf{H}(\textsf{sk})$.

The second is the signing algorithm $\textsf{EdDSA}.\textsf{sign}_{\textsf{sk}}(m)$ which takes a message $m \in \{0, 1\}^\ast$ and performs the following steps:
\vspace{5pt}
\begin{enumerate}
    \item Compute $k \in \mathbb{Z}_p$ by taking the first $\lambda$ bits of $\textsf{H}(\textsf{sk})$.
    \item Compute $r := \textsf{H}(h \mid\mid m) \in \mathbb{Z}_p$ where $h$ is the remaining $\lambda$ bits of $\textsf{H}(\textsf{sk})$ not used in step 1.
    \item Compute $R := r \cdot G \in E(\mathbb{Z}_p)$ and $s := r + \textsf{H}(R \mid\mid \textsf{pk} \mid\mid m) \cdot k \in \mathbb{Z}_p$.
    \item Return $\sigma := (R, s)$.
\end{enumerate}
\vspace{5pt}

The third is the verification algorithm $\textsf{EdDSA}.\textsf{vrfy}_{\textsf{pk}}(\sigma, m)$ which, given the signature $\sigma$ and the message $m$ and checks that $s \cdot G = R + \textsf{H}(R \mid\mid \textsf{pk} \mid\mid m) \cdot \textsf{pk}$. If this check is true, the algorithm returns \texttt{accept} else it returns \texttt{false}.

\subsection{Threshold Schemes}

First proposed by Shamir in 1979 \cite{shamir1979}, ($t$, $n$)-threshold schemes allow a set of $n$ participants to share a secret such that any $t$ out of the $n$ participants are required to cooperate to recover that secret. Any fewer than $t$ participants should not be able to recover any information about the secret.
\vspace{5pt}

\noindent\textsc{Shamir Secret Sharing:}\hspace{5pt} Many threshold schemes are based upon Shamir's original secret sharing scheme based on Lagrange interpolation \cite{shamir1979}. In a  ($t + 1$, $n$)-secret sharing scheme a secret $s \in \mathbb{Z}_q$ is shared among $P_1, \ldots, P_n$ participants through two algorithms. The first is $\textsf{share}(t, n)$ which uniformly generates a element $s \in \mathbb{Z}_q$ and elements $z_1, \ldots, z_t \in \mathbb{Z}_q$. and forms the polynomial 
\[
    f(x) = s + z_1x + \cdots + z_tx^t \in \mathbb{Z}_q[x]. 
\]
Each participant $P_j$ can be given a share $\mu_j := f(j) \in \mathbb{Z}_q$. The second is $\textsf{recover}(\mu_1, \ldots, \mu_{t + 1})$ which takes $t + 1$ shares and recovers $s$ through polynomial interpolation. Any coalition of participants running $\textsf{recover}$ with fewer than $t + 1$ shares can learn nothing about $s$. This method requires a trusted dealer to generate, distribute and delete the shares.
\vspace{5pt}

\noindent\textsc{Verifiable Secret Sharing:}\hspace{5pt} Verifiable Secret Sharing (VSS) gives a way to share a secret between participants $P_1, \ldots, P_n$ and verify that they are distributed the correct shares by the dealer. Feldman \cite{feldman1987} proposed a VSS scheme using Shamir's secret sharing combined with any homomorphic commitment scheme.

In the Feldman scheme, the dealer takes the polynomial $f$ as defined in $\textsf{share}$ and computes a polynomial commitment vector $\mathbf{c} := (g^s, g^{z_1}, \ldots, g^{z_t}) \in G^{t + 1}$ where $g$ is a group element in the group $G$. The commitment $\mathbf{c}$ is then sent to participants $P_1, \ldots, P_n$ and each participant is dealt a share $\mu_j := f(j) \in \mathbb{Z}_q$. A participant $P_j$ can use their share $\mu_j$ and the commitment $\mathbf{c}$ to check their share is correct by checking that $\prod^{t-1}_{i=0} c_i^{j^i} = g^{\mu_i}$. Even if the dealer is corrupted, the secret can be reconstructed if at least $t + 1$ parties received the correct shares.

In the original paper \cite{feldman1987}, Feldman does not specify actions to take if a participant detects an incorrect share. Pederson \cite{pederson1991a} designed a protocol to deal with the incorrect shares sent by a corrupted dealer in the Feldman VSS scheme. Furthermore, the Feldman scheme doesn't completely hide the secret $s$, since $c_0 = g^s$. Some applications may accept this, but others might require stronger guarantees.

Pederson's VSS scheme generates an additional polynomial $f' \in \mathbb{Z}_q[x]$ of degree at most $t$. The shares for each participant $P_j$ are now the pair $(\mu_j, \mu'_j) := (f(j), f'(j))$. As in the Feldman scheme the dealer sends the polynomial commitment $\mathbf{c} \in G^{t + 1}$ to $P_1,\ldots,P_n$, however $\mathbf{c}$ is now defined as the vector $(g^s h^s, g^{z_1} h^{z_1}, \ldots, g^{z_t} h^{z_t})$ where $h \in G$ is a random generator defined as a public parameter of the scheme. Each participant $P_j$ can then check its share is correct by checking that $\prod^{t-1}_{i=0}c_i^{j^i} = g^{\mu_j}h^{\mu'_j}$. As defined in \cite{pederson1991a}, if this check fails for any participant, they can raise an issue, and the protocol will terminate. This differs from how Feldman originally presented their scheme in \cite{feldman1987}, as it was assumed that an honest majority could recover the secret if a dishonest participant raised an issue.
\vspace{5pt}

\noindent\textsc{Asynchronous Verifiable Secret Sharing:}\hspace{5pt} The primary pitfall of the Feldman and Pederson VSS schemes outlined above is their inability to function correctly in asynchronous communication models. The notion of an Asynchronous VSS (AVSS) was first proposed by Ben-Or \emph{et al.} \cite{ben-or1994}, but the first protocol was outlined in work by Cachin \emph{et al.} \cite{cachin2002}.

In their AVSS protocol, Cachin \emph{et al.} uses a similar construction to that of Pederson \cite{pederson1991a}, with improved efficiency for asynchronous systems. Rather than a simple uni-variate polynomial, the dealer produces a two-dimensional sharing of a secret by generating a bi-variate polynomial $f \in \mathbb{Z}_q[x,y]$ with degree $k - 1$ and $f(0, 0) = s$. The dealer then produces a polynomial commitment by using a second random polynomial $f' \in \mathbb{Z}_q[x,y]$ to compute the matrix $\mathbf{C} = (\mathbf{c}_1, \ldots, \mathbf{c}_{k-1})$ where $\mathbf{c}_j = g^{\mu_j} h^{\mu'_j}$. The dealer then sends to participants $P_1, \ldots, P_n$ the commitment matrix $\mathbf{C}$ and the two share polynomials $a_j(y) = f(j, y)$ and $a'_j(y) = f'(j, y)$ and the two sub-share polynomials $b_j(x) = f(x, j)$ and $b'_j = f'(x, j)$.

When the participants receive their messages from the dealer, they send the points in which their share and sub-share polynomials overlap. When the participant receives messages that agree on $\mathbf{C}$ and contain valid points, they interpolate their share and sub-share polynomials from the received points using Lagrange interpolation. In the case of an honest dealer, the resulting polynomials are the same as those originally sent to them. The participants can then message the dealer that the sharing process is complete.

Despite its guarantees, the AVSS protocol defined by Cachin \emph{et al.} \cite{cachin2002} may not be as efficient as other options and has a communication complexity of $O(n^3\lambda)$ where $\lambda$ is the security parameter. Backes \emph{et al.} \cite{backes2013} proposed an AVSS protocol with communication complexity $O(n^2\lambda)$, which uses polynomial commitments that require their group supports a pairing operation, making it unusable for several commonly used digital signature schemes. AlHaddad \emph{et al.} \cite{alhaddah2021} proposed a protocol with communication complexity $O(n^2 \log n \cdot \lambda)$ which does not require pairing but relies on Bulletproofs \cite{bunz2018}. Finally, Groth and Shoup designed an AVSS scheme to work with threshold ECDSA, which, unlike the protocol defined by Cachin \emph{et al.}, only achieves computational privacy but achieves $O(n^2 \lambda)$ communication complexity.

\subsection{Distributed Key Generation}

The unfortunate fact of threshold cryptography schemes that use secret sharing techniques, such as Shamir's secret sharing, is their reliance on a trusted dealer. Distributed Key Generation (DKG) ensures that each protocol participant equally contributes to generating a shared secret. The core idea of a DKG protocol is that each participant uses a secret sharing protocol to disseminate a secret value. The participants then must reach a consensus on which secret values have been correctly shared. The resulting disseminated secrets can then be combined, with the outcome being a threshold private-public key pair.

Pederson \cite{pederson1991b} was the first to propose a practical two-round DKG scheme in which each participant acts as a dealer of the Feldman VSS protocol \cite{feldman1987}. Essentially, for participants $P_1, \ldots, P_n$ each participant $P_j$ generates a polynomial $f_j$ and a polynomial commitment $\mathbf{c}_j$ as in the Feldman VSS protocol and broadcasts $\mathbf{c}_j$ to the network. Participant $P_j$ then privately sends participants $P_\ell$ the share $\mu_{j\ell} = f_j(\ell)$ for $l = 1, \ldots, n$ and keeps $\mu_{jj}$ for itself. Each participant $P_j$ then verifies their received shares are consistent with their published commitments and computes their share as $\mu_j = \sum^n_{i=1} \mu_{ij}$.

Work by Gennaro \emph{et al.} \cite{gennaro2003} discovered a weakness in the Pederson DKG scheme \cite{pederson1991b} in which misbehaving participants can directly bias the distribution of secrets through issuing complaints against participants after seeing their sent secret. This resulted in the disqualification of the targeted nodes from contributing to the DKG protocol. Furthermore, Gennaro \emph{et al.} showed that Pederson's DKG scheme is secure in specific contexts, particularly more significant distributed systems where the chance for bias from the misbehaving participants is much smaller.

In 2007, Gennaro \emph{et al.} \cite{gennaro2007} proposed some modifications to the Pederson DKG scheme \cite{pederson1991b} to ensure its security properties are maintained even in smaller distributed systems. The authors introduce the notion of using the Feldman VSS \cite{feldman1987} and Pederson's own VSS protocol \cite{pederson1991a}, making the protocol require three rounds of communication. Furthermore, Gennaro \emph{et al.} added a commitment round, which forces the participants to perform the commitment round before revealing their inputs.

\section{Leaderless Consensus Through Distributed Signing}

This section illustrates achieving a byzantine fault-tolerant leaderless consensus using threshold signatures. Unlike signatures in a single-party setting, threshold signature schemes require the cooperation of $n$ participants up to a threshold $t$, each of which shares a secret that acts as a private key distributed across them. If fewer than $t$ of the participants are corrupted and act maliciously in the signing process, the signing will fail, resulting in no consensus among the participants. In this paper, an instantiation of the Flexible Round-Optimized Schnorr Threshold (FROST) signature scheme proposed by Komlo and Goldberg \cite{komlo2021} is used to achieve this.

As was discussed previously, many threshold schemes provide the notion of robustness \cite{gennaro2003}, whereby, if one participant acts maliciously, the remaining honest participants can detect that malicious action and complete the protocol as long as there is a threshold of at least $t$ honest participants. The FROST protocol trades robustness in favour of efficiency by allowing honest parties to identify malicious participants and aborting the protocol. Furthermore, in the instantiation of the FROST protocol given by Komlo and Golberg in their original paper \cite{komlo2021}, the use of a signature aggregator resulted in a semi-trusted distributed system, in which the aggregator could cause a denial of service. In this paper, rather than use the Schnorr signature scheme, the instantiation of FROST proposed will use the RFC 8032 standardised EdDSA signature scheme \cite{josefsson2017} due to its efficiency. Komlo and Goldberg briefly discuss using EdDSA in their original work \cite{komlo2021}.

\subsection{Trust Overlays}

\begin{figure}[ht]
\centering
\includegraphics[width=\linewidth]{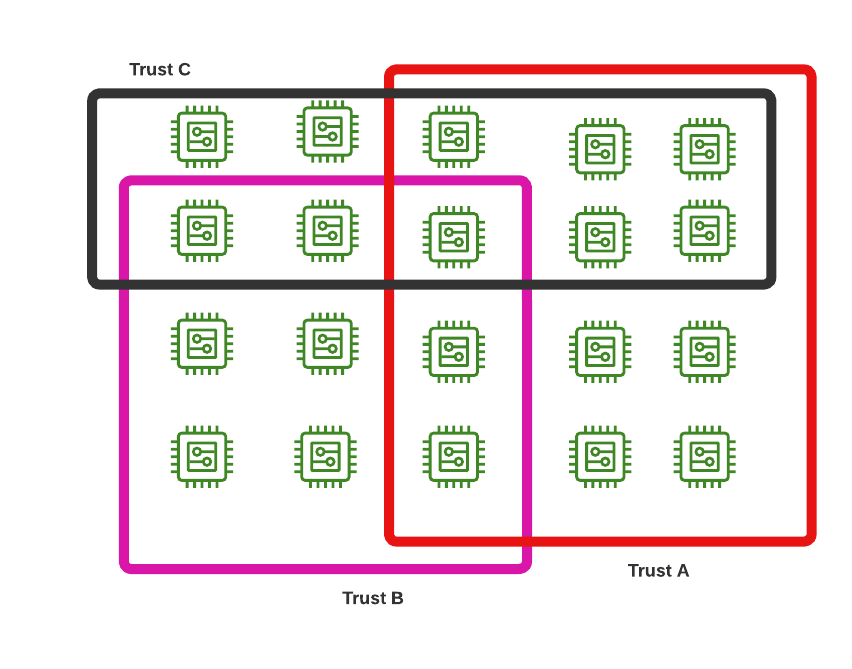}
\caption{Trust distribution}
\label{fig:trust01}
\end{figure}

Within critical infrastructure applications, we may need to overlap trust domains and where devices form a distributed grouping. As showin in Figure \ref{fig:trust01}, there are multiple nodes connected across three trust domains (Group A, Group B and Group C). The devices become self-organising and leaderless within each group, but the EdDSA public key for each group can be generated for the nodes in that specific group. In this way, none of the nodes in each group has to store a private key for the group but will only have fragments of the key, which can be used as a consensus to create a digital signature. The threshold method used can support Byzantine fault tolerance for devices becoming malicious or when specific devices become inoperable. By examining any of the nodes on the network, it will not be possible to discover the private key used for the group.

\subsection{Key Generation}

The generation of the secret key shares is built on top of the Pederson DKG scheme \cite{pederson1991b}; the details for this are given in Protocol 1. Pederson's original DKG was a straightforward process in which each participant acted as a dealer executing the Feldman VSS protocol \cite{feldman1987} in parallel.  Each participant would then derive their secret share as a sum of all shares received from each participant.  In addition to the Pederson DKG protocol, FROST requires that each participant $P_j$ generate proof of knowledge of their initial secret $s_j$ and provide that proof to all participants to verify.

The key generation protocol assumes the participants are formed using an implementation-defined mechanism. After participating in the Pederson DKG protocol, each participant $P_j$ should hold a value $\textsf{sk}_j$ that is the secret share they can use to participate in the signing protocol.  The other participants in the system can use the participant's public key share $\textsf{pk}_j$ to verify the correctness of $P_j$'s signature shares. The group public key $\textsf{pk}$ can be used by anyone external to the system to verify signatures issued by the participants in the future.

\begin{mdframed}[linewidth=1pt, frametitle={\normalfont \textsc{Protocol 1: Distributed Key Generation}}]
\noindent\textbf{Round 1}
\vspace{3pt}

\noindent 1  Each participant $P_j$ generates a degree $t - 1$ polynomial $f_j(x) := s_j + \sum^{t- 1}_{i = 1}a_{ji}x^i$ where $s_j$ is their secret.
\vspace{3pt}

\noindent 2  Each $P_j$ computes the proof of knowledge of their secret $s_j$ by $\pi_j := (G \cdot k, k + s \cdot \textsf{H}(j, \textsf{crs}, g \cdot s_j, G \cdot k))$ where $k \in \mathbb{Z}_q$ is uniformly sampled, and $\textsf{crs}$ is a common reference string used to prevent replay attacks.
\vspace{3pt}

\noindent 3  Each $P_j$ computes a public polynomial commitment $\mathbf{c}_j := (G \cdot s_j, G \cdot a_{j2}, \ldots, G \cdot a_{j(t-1)})$.
\vspace{3pt}

\noindent 4  Each $P_j$ broadcasts both $\mathbf{c}_j$ and $\pi_j$ to all other participants.
\vspace{3pt}

\noindent 5  Participant $P_J$ upon receiving $\mathbf{c}_\ell$ and $\pi_\ell$ for $\ell = 1, \ldots, n$ verifies that $\pi_{\ell0} = \mathbf{c}_{\ell0} \cdot -\textsf{H}(\ell, \textsf{crs}, G \cdot \mathbf{c}_{\ell0},\pi_{\ell0})$ and delete all proofs $\pi_\ell$. The protocol aborts on failure.
\vspace{6pt}

\noindent\textbf{Round 2}
\vspace{3pt}

\noindent 1  Each $P_j$ privately sends all participants $P_1, \ldots, P_n$ a share $\mu_j := f_j(j)$.
\vspace{3pt}

\noindent 2  Each $P_j$ verifies the share $\mu_\ell$ using the polynomial commitment $\mathbf{c}_\ell$ by checking that $G \cdot \mu_\ell = \sum^{t - 1}_{i = 0}\mathbf{c}_{\ell i}^{j^k \bmod q}$ for $\ell = 1, \ldots, n$. The protocol aborts on failure.
\vspace{3pt}

\noindent 3  Each $P_j$ computes their private signing share $\textsf{sk}_j := \sum^{n}_{i = 1}\mu_i$ and deletes each $\mu_\ell$ for $\ell = 1, \ldots, n$.
\vspace{3pt}

\noindent 4  Each $P_j$ computes their public verification share $\textsf{pk}_j := G \cdot \textsf{sk}_j$. The public key for the distributed system is $\textsf{pk} = \sum^n_{i = 1}\mathbf{c}_{i0}$
\end{mdframed}

\subsection{Threshold Signing}

The signature operation of the FROST protocol uses the shared secrets with secret conversion to non-interactively generate nonce values for each signature. Furthermore, the operation employs a binding technique to avoid forgery attacks without limiting the concurrency of the protocol.  In their original work, Komlo and Goldberg separate the preprocessing phase from the actual signing phase \cite{komlo2021}.  In this work, the two are combined into a two-round signing protocol.  Furthermore, the operation is described using the RFC 8032 standardised EdDSA signature algorithm \cite{josefsson2017} to work with the DKG operation described in Protocol 1.

The binding designed into the original FROST protocol \cite{komlo2021} prevents malicious participants from manipulating the resulting challenge $c$ for a set of participants performing a group signature operation. Each participant's response is bound to a specific message and the set of participants' commitments during that operation. The advantage of this additional process is that a combination of responses over different messages or participant commitment pairs results in an invalid signature.

\begin{mdframed}[linewidth=1pt, frametitle={\normalfont \textsc{Protocol 2: Signing}}]
\noindent\textbf{Round 1}
\vspace{3pt}

\noindent 1  Each participant $P_j$ generates an empty list $N_j$ then for $\ell = 1,\ldots,m$, uniformly sample two single-use nonces $(a_{j\ell}, b_{j\ell}) \in \mathbb{Z}^\ast_q \times \mathbb{Z}^\ast_q$ and compute the commitment shares $(A_{j\ell}, B_{j\ell}) := (G \cdot a_{j\ell}, G \cdot b_{j\ell})$ and append the shares to the list.
\vspace{3pt}

\noindent 2  Publish $(j, N_j)$, allowing all participants to access the commitment shares.
\vspace{6pt}

\noindent\textbf{Round 2}
\vspace{3pt}

\noindent 1  Each $P_j$ takes the nonce commitments $N_j$, constructs $\mathbf{n} := ((A_1, B_1), \ldots, (A_n, B_n))$ and checks that $A_\ell, B_\ell \in E(\mathbb{Z}_q)$ for $\ell = 1, \ldots, n$. The protocol aborts on failure.
\vspace{3pt}

\noindent 2  Each $P_j$ takes the message $m \in \{0, 1\}^\ast$ and computes a set of binding values $\beta_\ell := \textsf{H}_1(\ell \mid\mid m \mid\mid \mathbf{n})$ for $\ell = 1, \ldots, n$, the group commitment $R_j := \sum^n_{i = 0} A_i + B_i \cdot \beta_i$ and the challenge $c_j := \textsf{H}_2(R_j \mid\mid \textsf{pk} \mid\mid m)$.
\vspace{3pt}

\noindent 3  Each $P_j$ then computes $z_j := a_j + (b_j \cdot \beta_j) + \lambda_j \cdot \textsf{sk}_j \cdot c_j$ where $\lambda_j$ is the $j$\textsuperscript{th} Lagrange coefficient. $z_j$ can now be broadcast to the network.
\vspace{3pt}

\noindent 4  Each $P_j$ now acts as an aggregator \emph{via} the following steps:
\vspace{3pt}

\noindent 4.a. Compute $\beta_\ell := \textsf{H}_1(\ell \mid\mid m \mid\mid \mathbf{n})$ and $R_\ell := A_{\ell j} + B_{\ell j} \cdot \beta_\ell$ for $\ell = 1, \ldots, n$.
\vspace{3pt}

\noindent 4.b. Compute the values $R := \sum^n_{i = 1} R_i$ and $c := \textsf{H}_2(R \mid\mid \textsf{pk} \mid\mid m)$.
\vspace{3pt}

\noindent 4.c  Verify that $G \cdot z_\ell = R_\ell \cdot \textsf{pk}_\ell \cdot c \cdot \lambda_\ell$ for $\ell = 1, \ldots, n$. The protocol aborts on failure.
\vspace{3pt}

\noindent 4.d. Compute the systems response $z = \sum^n_{i = 1} z_i$ and publish the signature $\sigma := (R, z)$ and message $m$.
\end{mdframed}

\section{Signature Aggregation}

In their original paper, Komlo and Goldberg \cite{komlo2021} proposed FROST be instantiated in two ways: using a signature aggregator to reduce overhead or using broadcast messages to aggregate the signatures, incurring significantly more communication overhead.  To avoid $O(n^2\lambda)$ verification work, a gossip protocol is designed to speed up signature aggregation without electing a signature aggregator. The main observation is that any participant who has verified several responses can aggregate them together into a single response and forward this to the next participant.  Through careful use of the gossip protocol and aggregation, the verification time can be decreased significantly, especially considering large distributed systems with many individual aggregators.

The gossip protocol outlined in Protocol 3 has each party send its currently aggregated DKG transcript to $O(c \log n)$ participants and terminates when it has reached an agreement on a full transcript, in other words, when the transcript has reached $t + 1$ contributions.  In this case, $c$ is a small success parameter such that $c \geq 4$.  Deciding when to terminate the protocol is difficult since the full aggregated transcripts may all be different.  This is why the broadcast must be invoked to ensure that the participants know which transcript is to be used.  However, the goal was to reduce broadcasts achieved by making the broadcast only happen with a probability of $2/n$ in any given round of the gossip protocol which makes the protocol likely to terminate in $O(c\log n)$ rounds of messaging.  Participants can then use any implementation-defined convention to agree on which transcript to use.  The resulting communication complexity is $O(c\log^2 n)$ broadcasts. 

\begin{mdframed}[linewidth=1pt, frametitle={\normalfont \textsc{Protocol 3: Aggregation Gossip Protocol}}]

\noindent 1  Each $P_j$ sends $(z_j, \textsf{transcript}_j)$ to a random $2/n$ of the participants $P_1, \ldots, P_n$.
\vspace{3pt}

\noindent 2  If $P_j$'s transcript $\textsf{transcript}_j$ contains $t + 1$ contributions broadcast the transcript and end the protocol with probability $2/n$.
\vspace{3pt}

\noindent 2  If a send request is received by $P_j$ from $P_\ell$ for $\ell = 1, \ldots, n$ verify that $z_\ell$ and $\textsf{tanscript}_\ell$ are correct then set $\textsf{transcript}_j := \textsf{transcript}_j + \textsf{transcript}_\ell$ and repeat the protocol.

\end{mdframed}

\section{Implementation and Evaluation}

In this section, we measured the communication overhead of the DKG of the proposed framework as a function of the network size, where the number of participating nodes ranges from a minimum of four nodes to a maximum of 255 nodes.  The core focus of the evaluation is the usage of the FROST method and where   each node can check that they have a valid share of the distributed private key and also whether other nodes have valid shares; each group then has an associated public key. 

In order to understand the basic dynamics of the systems, a prototype of the system was built using the Kryptology library \cite{asecuritysite_19437} and validated for a distributed EdDSA signature for a $t$ out of $n$ scheme. Other demonstrators were created that used the GG20 \cite{gennaro2020one,asecuritysite_87116} and DKLS \cite{doerner2019threshold,asecuritysite_56432} distributed signature schemes. Both GG20 and DKLS use ECDSA signatures and perform worse in scaling the network. Furthermore, the original GG20 protocol has been shown to have security weaknesses and is not advised for current implementation (see CVE-2023-33241 \cite{CVE-2023-33241}). The FROST method has three main phases: secret key generation, verification key generation, and splitting secret keys and rebuilding the verification key. Table \ref{tab:1} shows the results running on a t2.medium instance in AWS (two vCPUs and 4GB of memory) for the broadcast and verification phase of Round 1. The broadcast period is acceptable, but the time penalty for the verification for 128 nodes is over 7.4 seconds, and for 255 nodes, it is nearly 31 seconds. The table also shows that the signing phase in the system performs well, with the time taken for signing remaining consistently under 10.3 milliseconds under various numbers of nodes in the 3-from-n threshold scheme, indicating that the computational overhead is fairly constant in creating the signature.

\begin{table}[ht]
    \centering
    \caption{Results for Protocol 1 (Distributed Key Generation).}
    \label{tab:1}
    \begin{tabular}{ccll}
        \toprule
        & & \multicolumn{2}{c}{Time (ms)} \\
        \cmidrule(lr){3-4}
        Threshold ($t$) & Participants ($n$) & Round 1 & Round 2 \\ 
        \midrule
        3 & 4 & 3.4 & 12.4 \\ 
        3 & 8 & 4.0 & 30.1 \\ 
        3 & 16 & 7.0 & 122.6 \\ 
        3 & 32 & 17.5 & 442.1 \\ 
        3 & 64 & 33.2 & 2,030 \\ 
        3 & 128 & 71.9 & 7,460 \\ 
        3 & 255 & 222.1 & 30,760 \\
        \bottomrule
    \end{tabular}
\end{table}

\section{Conclusion}

This paper outlines an EdDSA-based framework to create leaderless trusted groups within a large-scale network, which can be applied to the protection of critical infrastructure. The proposed approach is found to outperform ECDSA-based approaches in terms of scalability and could have great benefits where an EVM (Ethereum Virtual Machine) is not required. The approach defined is also able to act autonomously from other networked systems. The evaluation results have shown that the greatest computational overhead of the proposed approach is in the second round of the DKG, where it can take several seconds for the distributed key to be verified. Once the keys are distributed, the computational resources required for the subsequent signing process remain relatively constant, even when the network is scaled up to include as many as 255 nodes.

\bibliographystyle{IEEEtran}
\bibliography{references}

\end{document}